\documentclass[twoside,11pt]{article}

\usepackage{blindtext}

\usepackage{jmlr2e}
\usepackage{algorithmic}
\usepackage{graphicx}
\usepackage{hyperref}
\usepackage{xcolor}
\usepackage{algorithm}

\newcommand{\libname}[0]{marl-jax}

\usepackage{lastpage}

\firstpageno{1}

\begin{document}

\title{\libname: Multi-agent Reinforcement Leaning Framework}

\author{\name Kinal Mehta$^{1}$ \email kinal.mehta@research.iiit.ac.in \\
       \name Anuj Mahajan$^{2}$ \email anuj.mahajan@cs.ox.ac.uk \\
       \name Pawan Kumar$^{1}$ \email pawan.kumar@iiit.ac.in \\
       \addr $^{1}$International Institute of Information Technology Hyderabad, India \\
       \addr $^{2}$University of Oxford, Oxford, UK
}

\editor{}

\maketitle

\begin{abstract}
Recent advances in Reinforcement Learning (RL) have led to many exciting applications. These advancements have been driven by improvements in both algorithms and engineering, which have resulted in faster training of RL agents.
We present \libname, a multi-agent reinforcement learning software package for training and evaluating social generalization of the agents. 
The package is designed for training a population of agents in multi-agent environments and evaluating their ability to generalize to diverse background agents. It is built on top of DeepMind's JAX ecosystem~\cite{deepmind2020jax} and leverages the RL ecosystem developed by DeepMind. Our framework \libname~is capable of working in cooperative and competitive, simultaneous-acting environments with multiple agents. The package offers an intuitive and user-friendly command-line interface for training a population and evaluating its generalization capabilities.
In conclusion, \libname~provides a valuable resource for researchers interested in exploring social generalization in the context of MARL. The open-source code for \libname~is available at: \href{https://github.com/kinalmehta/marl-jax}{https://github.com/kinalmehta/marl-jax}
\end{abstract}


\section{Introduction}
Multi-agent reinforcement learning (MARL) is an important framework for training autonomous agents that operate in dynamic environments with multiple learning agents. Many potential real-world applications require the trained agents to cooperate with humans or agents not seen during training. That is, they should be able to zero-shot generalize to novel social partners. Most of the existing MARL frameworks~\cite{samvelyan19smac, papoudakis2021epymarl, sarkarPantheonRLMARLLibrary2022, malib, hu2022marllib} are either designed for cooperative MARL research or naively extend existing single-agent RL frameworks to work with multiple agents.

On the contrary, \libname~is designed specifically for multi-agent research and facilitate the training and assessment of the generalization capacities of multi-agent reinforcement learning (MARL) algorithms when facing new social partners. We utilize the functionalities of JAX~\cite{jax2018github} including autograd, vectorization through \emph{vmap}, parallel processing through \emph{pmap}, and compilation through \emph{jit}, resulting in highly optimized training for multiple agents.

\section{Related Works}
The RL community has developed several frameworks targeting various aspects such as implementation simplicity, ease of adaptation and scaling deep RL agents. In \libname, we focus on ease of experimentation and adaption for training a population of agents in multi-agent environments.

A number of libraries that concentrate on single-agent reinforcement learning have been created, such as stable-baselines3~\cite{raffinStableBaselines3ReliableReinforcement2021}, dopamine~\cite{castro18dopamine}, acme~\cite{hoffmanAcmeResearchFramework2022}, RLlib~\cite{liangRLlibAbstractionsDistributed2018}, and CleanRL~\cite{huangCleanRLHighqualitySinglefile2022}. These libraries prioritize features like modularity by providing useful abstractions, ease of use by requiring minimal code to get started, distributed training and ease of comprehension and reproducibility. Other libraries such as Reverb~\cite{cassirerReverbFrameworkExperience2021}, rlax~\cite{deepmind2020jax}, and launchpad~\cite{yangLaunchpadProgrammingModel2021} concentrate on specific components of an RL system.

For multi-agent reinforcement learning, several libraries have been developed, including PyMARL~\cite{samvelyan19smac}, epymarl~\cite{papoudakis2021epymarl}, RLlib~\cite{liangRLlibAbstractionsDistributed2018}, Mava~\cite{pretoriusMavaResearchFramework2021}, and PantheonRL~\cite{sarkarPantheonRLMARLLibrary2022}. RLlib and PantheonRL enhance existing single-agent RL algorithms to enable multi-agent training, while Mava, PyMARL, and epymarl are specifically designed for MARL but only support cooperative environments.

The advancements of Reinforcement Learning (RL) algorithms have been greatly influenced by libraries providing a range of environments. Single agent RL has been aided by libraries such as OpenAI Gym~\cite{2016openaigym} and dm-env~\cite{dm_env2019}, which established the framework for environment interactions. Multi-agent RL has been supported by SMAC~\cite{samvelyan19smac} and PettingZoo~\cite{terry2021pettingzoo}. Recently, DeepMind has contributed to the field by open-sourcing MeltingPot~\cite{agapiouMeltingPot2022}, a library for evaluating multi-agent generalization to new social partners at scale. Similarly, efforts for measuring generalization in cooperative multi-agent settings \cite{mahajan2022generalization} are being supported by libraries like \cite{ellis2022smacv2}.

Several recent works \cite{hoffmanAcmeResearchFramework2022, huangCleanRLHighqualitySinglefile2022, deepmind2020jax, pretoriusMavaResearchFramework2021} have begun utilizing JAX \cite{jax2018github} due to its various benefits. These benefits include auto-vectorization, just-in-time compilation, and easy multi-GPU scaling. 

\section{\libname}
Inspired by Acme~\cite{hoffmanAcmeResearchFramework2022}, we share a lot of design philosophies with it. Reverb~\cite{cassirerReverbFrameworkExperience2021} is used as a data-store server for the replay buffer. Launchpad~\cite{yangLaunchpadProgrammingModel2021} is used for distributed computing. We use JAX~\cite{jax2018github} as the numerical computation backend for neural networks. 
We use \emph{dm-env} API as our environment interaction API and extend it for multi-agent environments.

\subsection{System Architecture}

We implement four different training architectures

\subsubsection{Single Threaded}
\begin{figure}
    \centering
    \includegraphics[width=0.95\linewidth]{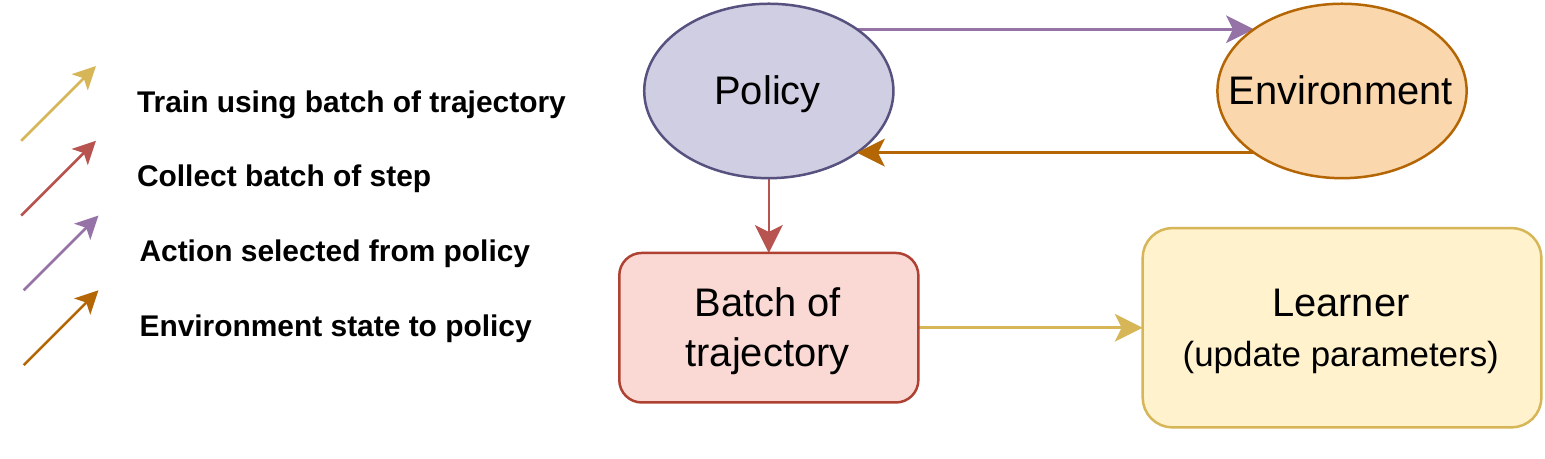}
    \caption{Single-threaded RL training pipeline}
    \label{fig:single_process}
\end{figure}

Figure~\ref{fig:single_process} illustrates the sequential flow of operations in a single-threaded RL training process. At each step, the agent receives an observation from the environment, selects an action based on its policy, interacts with the environment, and receives a reward. The observation, action, reward, and next observation are collected to form a  batch and then used to update agent's policy and value function using gradient descent. This process continues iteratively until the desired convergence or a specified number of iterations is reached. Algorithm \ref{algo:single-threaded} describes the pseudocode for training a policy in singe-threaded manner.

\begin{algorithm}
\caption{\label{actor} Single-threaded RL training}
\begin{algorithmic}[1] 
\label{algo:single-threaded}
\REPEAT
\STATE params = initialize params for the policy and value function
\STATE sequence = [\quad]
\STATE timestep = environment.reset()
\REPEAT
\STATE actions = predict\_actions(params, timestep.observation)
\STATE new\_timestep = environment.step(actions)
\STATE sequence.append((timestep, actions))
\STATE timestep = new\_timestep
\UNTIL{not timestep.last() \textbf{or} len(sequence) $<$ MAX\_BATCH\_SIZE}
\STATE params, logs = sgd\_step(params, sequence)
\STATE logger.write(logs)
\UNTIL{actor\_steps $<$ MAX\_STEPS}
\end{algorithmic}
\end{algorithm}

\subsubsection{Synchronous Distributed}
\begin{figure}
    \centering
    \includegraphics[width=0.95\linewidth]{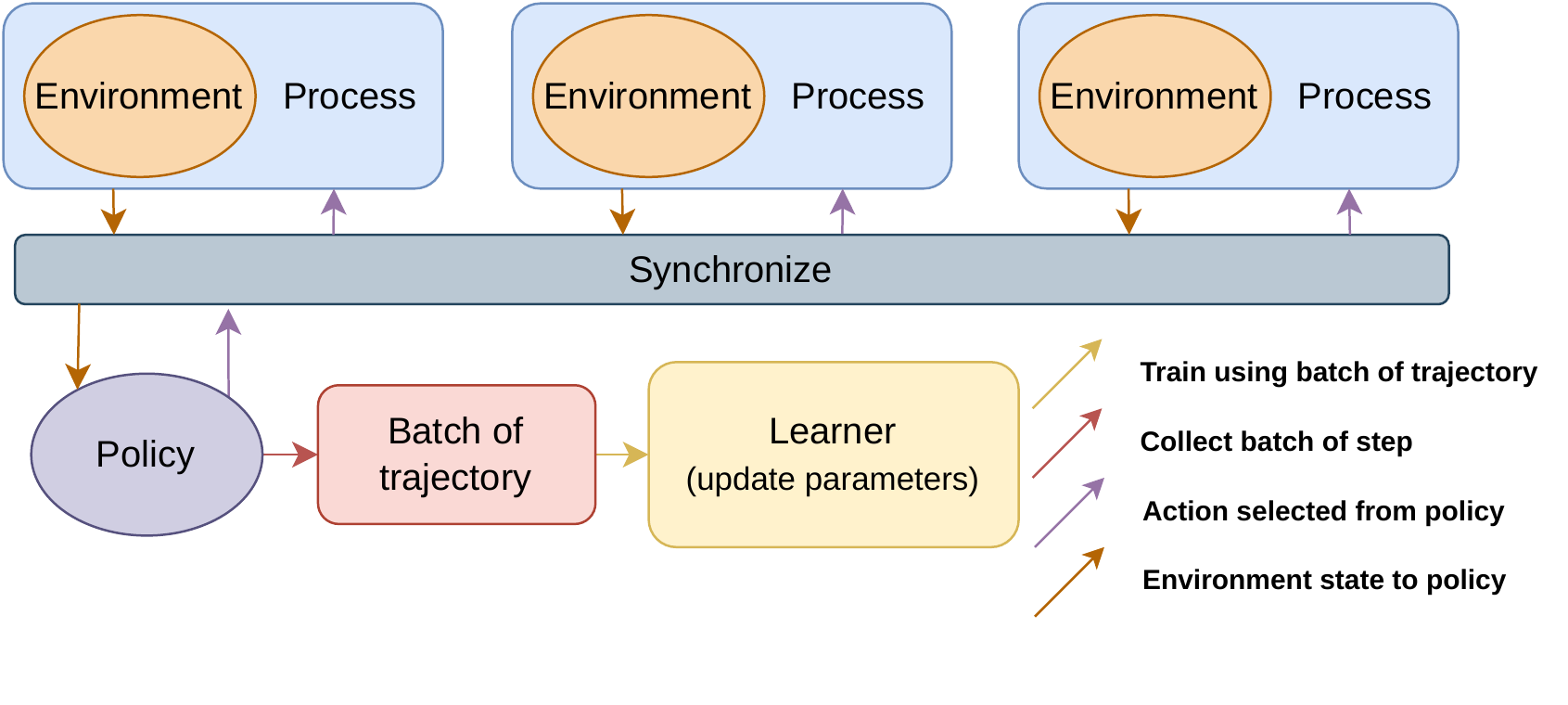}
    \caption{RL training with vectorized environment}
    \label{fig:sync_arch}
\end{figure}

The synchronous distributed architecture builds upon the single-threaded architecture by utilizing multiple environment instances running in parallel processes to collect a batch of experiences simultaneously. Each environment synchronously interacts with a common policy, generating sequences of states, actions, rewards, and next observations. This batch of sequences is used to update the policy and value function weights through gradient descent. By leveraging parallelization, the synchronous distributed architecture enables more efficient data collection and faster updates, leading to accelerated training and improved convergence in reinforcement learning. Fig.\ref{fig:sync_arch} illustrates how synchronous parallelization is achieved by running each environment instance in a separate process. The pseudocode is similar to that described in algorithm\ref{algo:single-threaded}, with the main difference being that each interaction with the environment results in a batch of data collected from environments across all processes.

\subsubsection{IMPALA-style Asynchronous Distributed}
\begin{figure}
    \centering
    \includegraphics[width=0.95\linewidth]{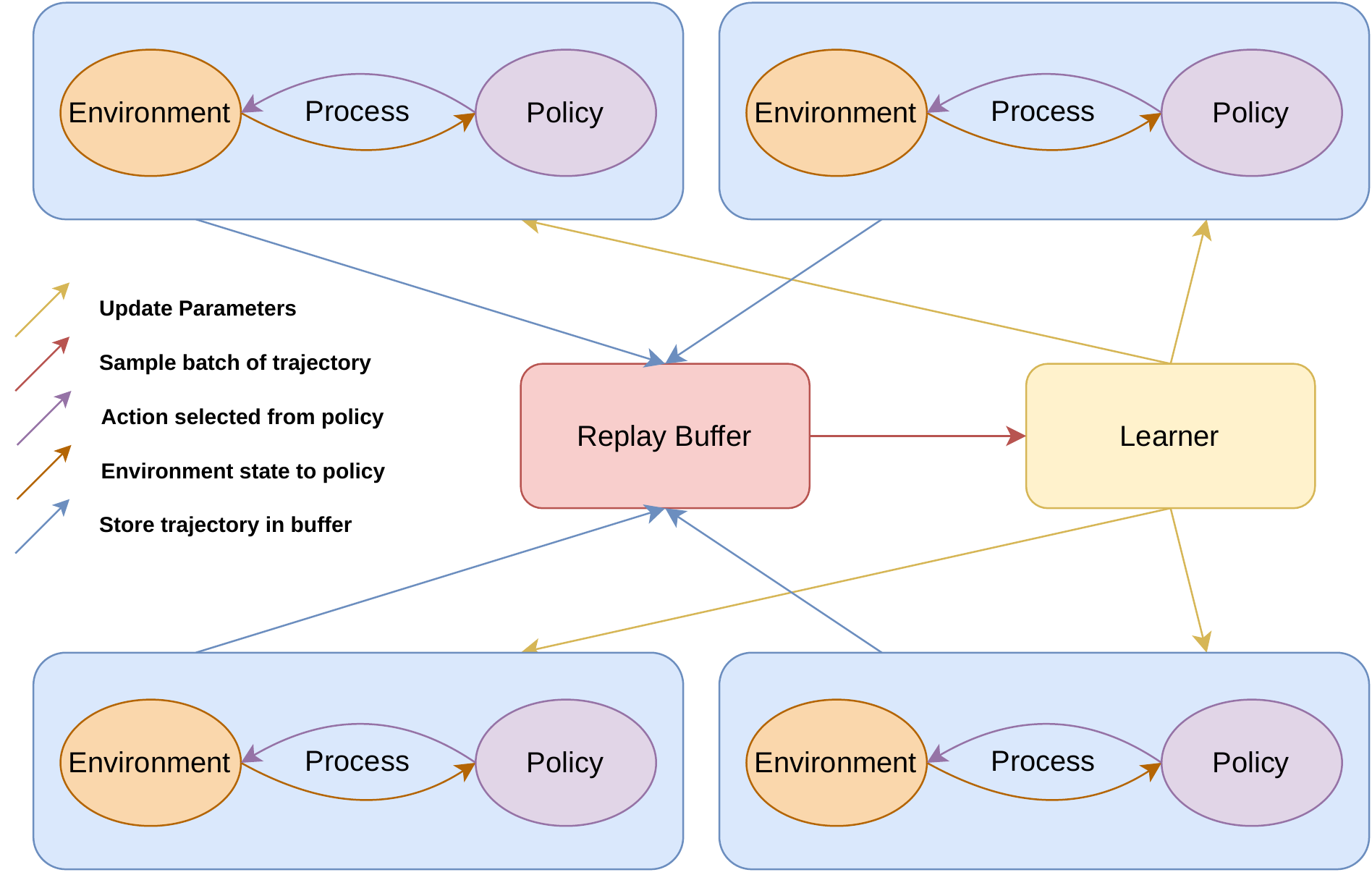}
    \caption{Asynchronous distributed RL training pipeline}
    \label{fig:impala_arch}
\end{figure}

In the IMPALA-style asynchronous distributed training architecture, multiple actors run in parallel and interact with their respective environments asynchronously. Each actor collects trajectories of experience by executing its own copy of the current policy. 
The three main components running in parallel as separate processes are described below
\begin{itemize}
    \item \textbf{Environment Loop:} The environment loop process interacts with the environment using the available policy and adds the collected experience to the replay buffer. Multiple parallel environment loop processes are run, each with its own copy of the environment and policy parameters. We use CPU inference for action selection on each process. To keep the policy parameters in sync with the learner process, the parameters are periodically fetched from the learner process. The action selection step is optimized using \emph{vamp} auto-vectorization to select the action for all agents in the environment. Algorithm~\ref{algo:async_actor} shows the pseudocode for this process.
    \item \textbf{Learner:} The actual policy learning happens in this process. The learner fetches experience from the replay buffer and performs the optimization step on policy and value function parameters. We use \emph{pmap} to auto-scale the optimization step to multiple GPUs and \emph{vmap} based auto-vectorization to perform the optimization step for all agents in parallel. Algorithm~\ref{algo:async_learner} shows the pseudocode for this process.
    \item \textbf{Replay Buffer:} A separate process with reverb~\cite{cassirerReverbFrameworkExperience2021} server is used as a replay buffer. All the actors add experience to this server, and the learner process samples experience from the server to optimize for policy and value function parameters.
\end{itemize}

Figure~\ref{fig:impala_arch} illustrates the how data flows between different components enabling asynchronous experience collection and training of the RL agent. 

\begin{algorithm}
\caption{\label{actor} Asynchronous Environment loop}
\begin{algorithmic}[1] 
\label{algo:async_actor}
\REPEAT
\STATE params = fetch latest params from learner
\STATE episode = [\quad]
\STATE timestep = environment.reset()
\REPEAT
\STATE actions = predict\_actions(params, timestep.observation)
\STATE new\_timestep = environment.step(actions)
\STATE episode.append((timestep, actions))
\STATE timestep = new\_timestep
\UNTIL{not timestep.last()}
\STATE replay\_buffer.write(episode)
\STATE logger.write(logs)
\UNTIL{actor\_steps $<$ MAX\_STEPS}
\end{algorithmic}
\end{algorithm}

\begin{algorithm}
\caption{\label{learner} Asynchronous Learner}
\begin{algorithmic}[1]
\label{algo:async_learner}
\REPEAT
\STATE batch = replay\_buffer.sample()
\STATE new\_params, logs = sgd\_step(params, batch)
\STATE logger.write(logs)
\UNTIL{actor\_steps $<$ MAX\_STEPS}
\end{algorithmic}
\end{algorithm}

\subsubsection{Sebulba: Asynchronous Distributed with Inference Server}
\begin{figure}
    \centering
    \includegraphics[width=0.95\linewidth]{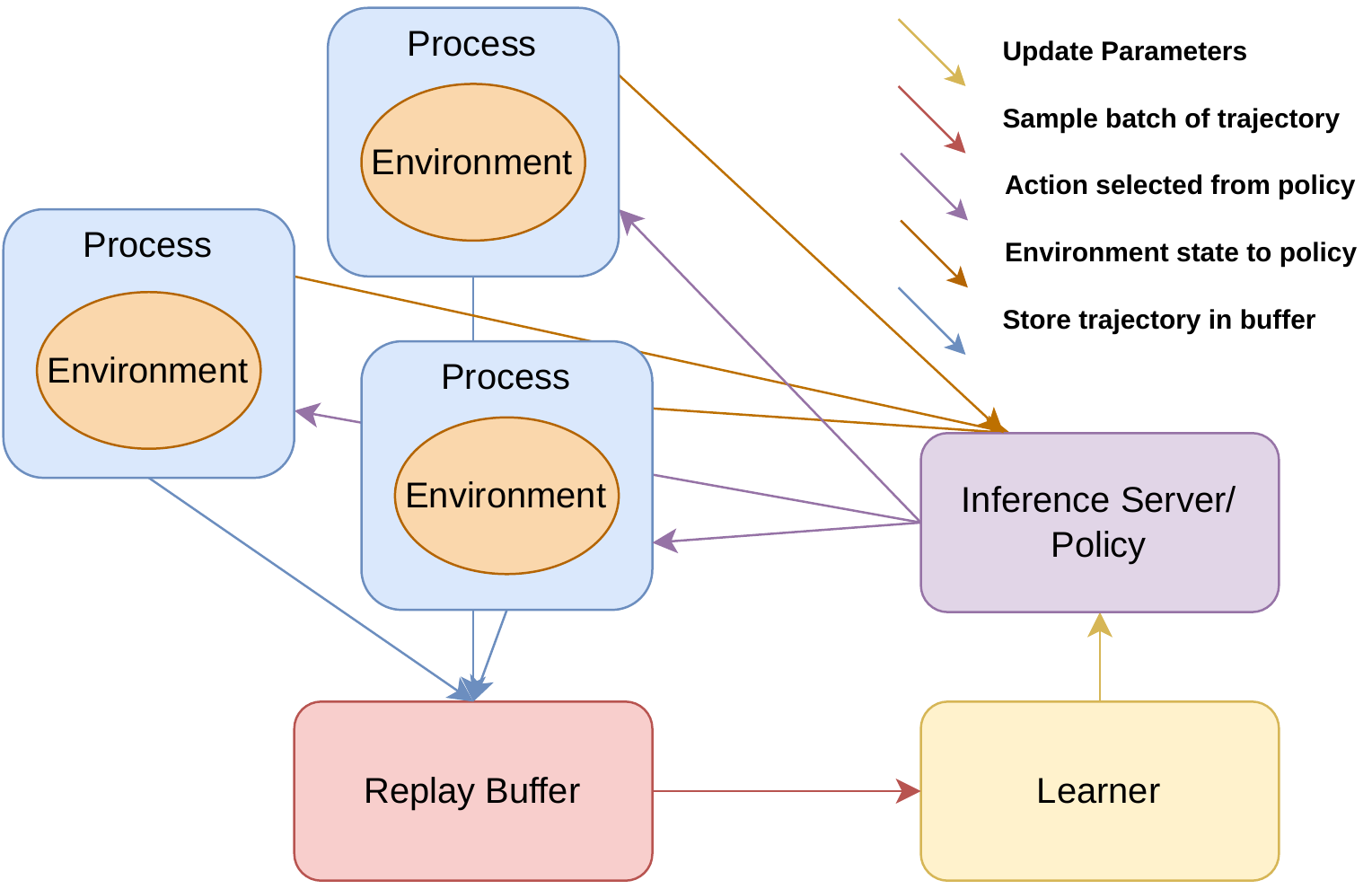}
    \caption{Asynchronous distributed with Inference Server}
    \label{fig:subalba_arch}
\end{figure}

Inspired by Sebulba architecture from Podracer~\cite{hessel2021podracer} and seed-rl~\cite{espeholt2019seed}, this architecture uses a common inference server in the asynchronous distributed architecture.
When a common inference server is used in the asynchronous distributed architecture for RL training, the architecture is further enhanced to centralize the inference process. In this setup, multiple actors interact with their respective environments asynchronously, collecting trajectories of experience as before. However, instead of each actor performing its own inference, they send their collected experiences to a common inference server which usually has access to hardware-accelerator such as GPU.

By utilizing a common inference server, several advantages can be achieved. First, it reduces the computational load on the individual actors, as they no longer need to perform their own inference. This enables the actors to focus on data collection, resulting in more efficient and faster interaction with the environment.
Second, the use of a shared policy network ensures that all actors are making decisions based on the same set of parameters. This improves the consistency and stability of the training process, as it prevents any discrepancies that may arise from differences in local copies of the policy network.

Figure~\ref{fig:subalba_arch} shows the block-diagram and data-flow in asynchronous architecture with inference server. The pseudocode for learner will be exactly the same as the Algorithm~\ref{algo:async_learner} and for actor it is same as Algorithm~\ref{algo:async_actor} where the \emph{predict\_actions} functions will be a call to the inference server instead of local policy network.

\subsection{Supported Environments}
We support two multi-agent environment suits, which consist of simultaneous acting homogeneous agents.

\subsubsection{Overcooked} 

The Overcooked environment~\cite{carrollUtilityLearningHumans2019overcooked} is a popular benchmark in the field of Multi-Agent Reinforcement Learning (MARL) that simulates a cooperative cooking scenario based on the popular game \emph{Overcooked}. It provides a challenging and interactive environment where multiple agents collaborate to prepare dishes in a virtual kitchen.

In Overcooked, the goal is to efficiently work together as a team to prepare and serve a variety of meals. The agents control different characters within the kitchen and must coordinate their actions to complete tasks such as chopping ingredients, cooking, and delivering finished dishes to customers. Collaboration and coordination are essential to maximize efficiency and achieve high scores.

The environment features various elements that add complexity to the task. For example, the kitchen layout may include obstacles that require agents to navigate around, limited resources like cutting boards and stoves that need to be shared, and time-sensitive customer orders that must be fulfilled promptly. Additionally, agents need to strategize and communicate effectively to optimize their actions and avoid potential bottlenecks or collisions.

Overcooked is designed to test the ability of MARL algorithms to solve cooperative tasks in dynamic and complex environments. It challenges agents to exhibit skills such as coordination, planning, communication, and adaptive decision-making.

\subsubsection{Melting Pot}

Melting Pot~\cite{agapiouMeltingPot2022} suite designed with the objective of evaluating generalization to novel situations and coplayers. The Melting Pot 2.0 suite consists of $50$ different environments and over $256$ unique test scenarios to evaluate the trained population of agents on broad range of topics such as social dilemmas, task partioning, resource sharing, etc

Melting Pot evaluation methodology is captured by the following equation:

\textbf{Substrate + Background Population = Scenario}
\begin{itemize}
    \item \textbf{Substrate:} The term "substrate" refers to the static or physical aspects of the environment in a simulation. It encompasses elements such as the layout of the map, the placement of objects, the rules governing their movement, and the physics involved. In essence, the substrate defines the stationary or unchanging components of the environment's dynamics. It sets the foundation and structure upon which other dynamic elements and interactions can take place. By defining the substrate, the simulation establishes the framework for how the environment behaves and provides a stable backdrop against which other entities and events can unfold.
    
    \item \textbf{Background Population:} The term "background population" refers to a group of simulated entities within a simulation that have their own agency or ability to take actions and make decisions. In other words, these entities are not passive or static; they actively participate in the simulation and contribute to its dynamics. They can interact with other entities, respond to stimuli or events, and potentially influence the overall behavior and outcomes of the simulation. The background population adds an element of realism and complexity to the simulation, making it more dynamic and reflective of real-world scenarios.
     
    \item \textbf{Scenario:} In the context of simulation or modeling, a scenario is created by combining the substrate and the background population. The substrate refers to the static or physical part of the environment, such as the layout, objects, and physics rules. On the other hand, the background population consists of simulated entities with agency, meaning they can take actions and make decisions within the simulation.

By integrating the substrate and the background population, a scenario is formed that represents a specific setting or situation within the simulation. The substrate provides the foundation, defining the physical attributes and constraints of the environment. This includes factors like the terrain, structures, objects, and their spatial arrangement. The substrate sets the stage for interactions and events to occur.

The background population adds a dynamic aspect to the scenario. These simulated entities have their own behaviors, goals, and decision-making processes. They can interact with each other, respond to stimuli or events in the environment, and potentially influence the overall dynamics of the scenario. The actions and interactions of the background population create a realistic and evolving simulation environment.

Together, the substrate and background population create a scenario that encapsulates a particular context or situation within the simulation. This scenario can be designed to simulate real-world scenarios, test hypotheses, study the behavior of complex systems, or provide a platform for experimentation and analysis. By carefully defining the substrate and background population, researchers and practitioners can create meaningful and informative scenarios that capture the intricacies of the system being studied.
\end{itemize}

\begin{figure}
    \centering
    \includegraphics[height=96pt]{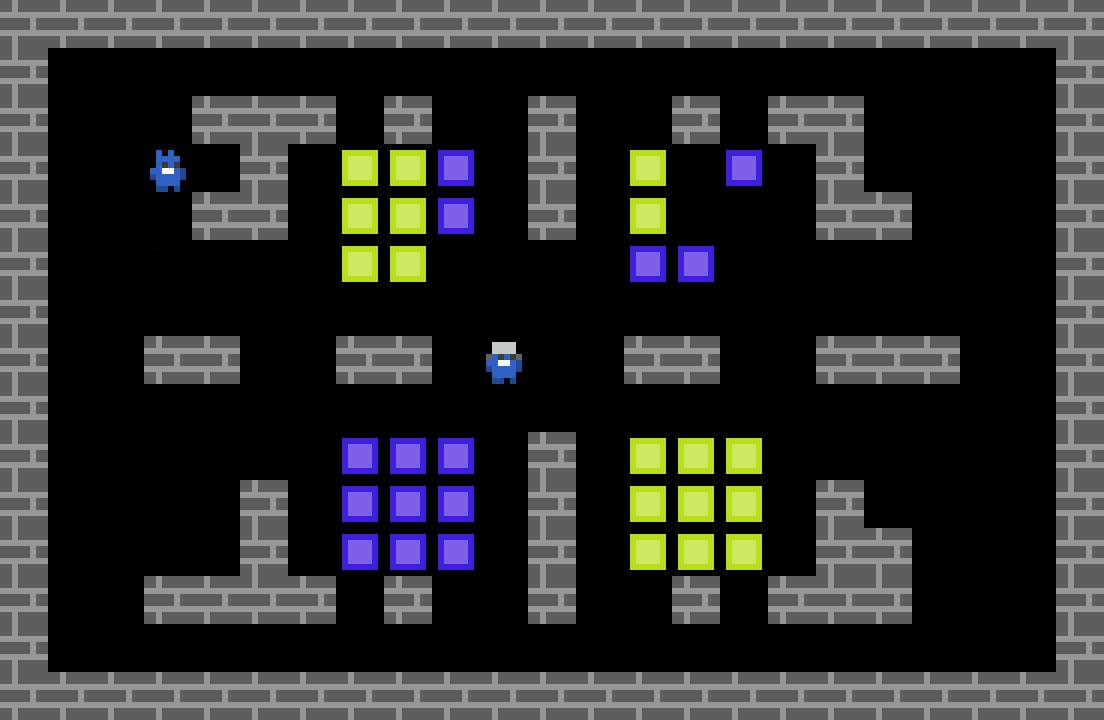}
    \includegraphics[width=0.5\textwidth]{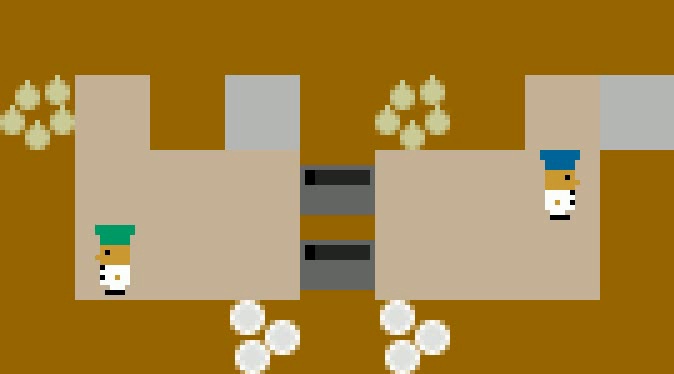}
    \caption{\libname~supports two major environment suits, Meltingpot~\cite{agapiouMeltingPot2022} and Overcooked~\cite{carrollUtilityLearningHumans2019overcooked}}
    \label{fig:env_supports}
\end{figure}

\subsection{Algorithms Implemented}
We currently support two major algorithms
\begin{itemize}
    \item \textbf{Actor-Critic Baseline:} A standard actor-critic based independent learning algorithm using V-trace~\cite{espeholtIMPALAScalableDistributed2018} for off-policy corrections.
    \item \textbf{OPRE:} Options as Responses~\cite{vezhnevetsOptionsResponsesGrounding2020} follows actor-critic based learning but its objective is specifically designed to generalize to novel partners. It is used as one of the baseline in MeltingPot~\cite{agapiouMeltingPot2022}. We are the first to provide an open-source implementation of OPRE.
\end{itemize}

\subsection{Utilities}
We provide two major utilities 1) \textit{train.py} and 2) \textit{evaluate.py}
\begin{itemize}
    \item \textbf{train.py:} The entry point for training a population of agents in the given environment
    \item \textbf{evaluate.py:} Used to evaluate the generalization performance on with various partner agents
    \item \textbf{evaluation\_results.py:} Aggregates the evaluation results by averaging across multiple seeds and presents a table.
\end{itemize}

\section{Results}

\begin{table}[ht]
   \begin{center}  
    \begin{tabular}{|c|c|c|c|}
    \hline
     & \bfseries OPRE & \bfseries IMPALA \\
    \hline
    \bfseries Substrate & 0.00 & 0.00 \\
    \bfseries Scenario 0 & 4.91 & -7.10 \\
    \bfseries Scenario 1 & 3.79 & -2.65 \\
    \bfseries Scenario 2 & 10.52 & 5.97 \\
    \bfseries Scenario 3 & 13.52 & 11.52 \\
    \bfseries Scenario 4 & 6.60 & 0.66 \\
    \hline
    \end{tabular}
    \caption{Evaluation on Running with Scissors in the matrix}
    \label{tab:rws}
    \end{center}
\end{table}

\begin{figure}
    \centering
    \includegraphics[width=0.6\linewidth]{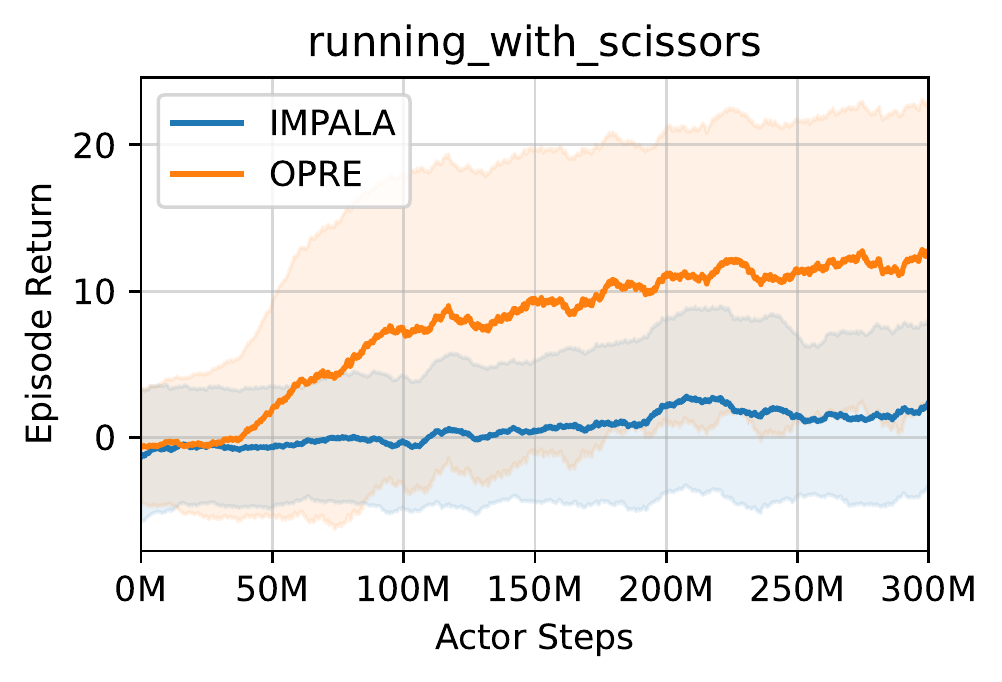}
    \caption{Training Plot on Running with Scissors in the matrix}
    \label{fig:rws}
\end{figure}

\begin{table}[ht]
   \begin{center}  
    \begin{tabular}{|c|c|c|c|}
    \hline
     & \bfseries IMPALA & \bfseries OPRE \\
    \hline
    \bfseries Substrate & 106.85 & 38.18 \\
    \bfseries Scenario 0 & 131.00 & 59.71 \\
    \bfseries Scenario 1 & 176.54 & 114.69 \\
    \bfseries Scenario 2 & 79.58 & 27.97 \\
    \bfseries Scenario 3 & 62.80 & 41.76 \\
    \bfseries Scenario 4 & 48.63 & 38.75 \\
    \bfseries Scenario 5 & 65.82 & 47.66 \\
    \bfseries Scenario 6 & 101.83 & 40.34 \\
    \bfseries Scenario 7 & 83.33 & 49.82 \\
    \bfseries Scenario 8 & 77.75 & 32.59 \\
    \bfseries Scenario 9 & 78.41 & 74.62 \\
    \hline
    \end{tabular}
    \caption{Evaluation on Prisoners Dilemma in the Matrix}
    \label{tab:dilemma}
    \end{center}
\end{table}

\begin{figure}
    \centering
    \includegraphics[width=0.7\linewidth]{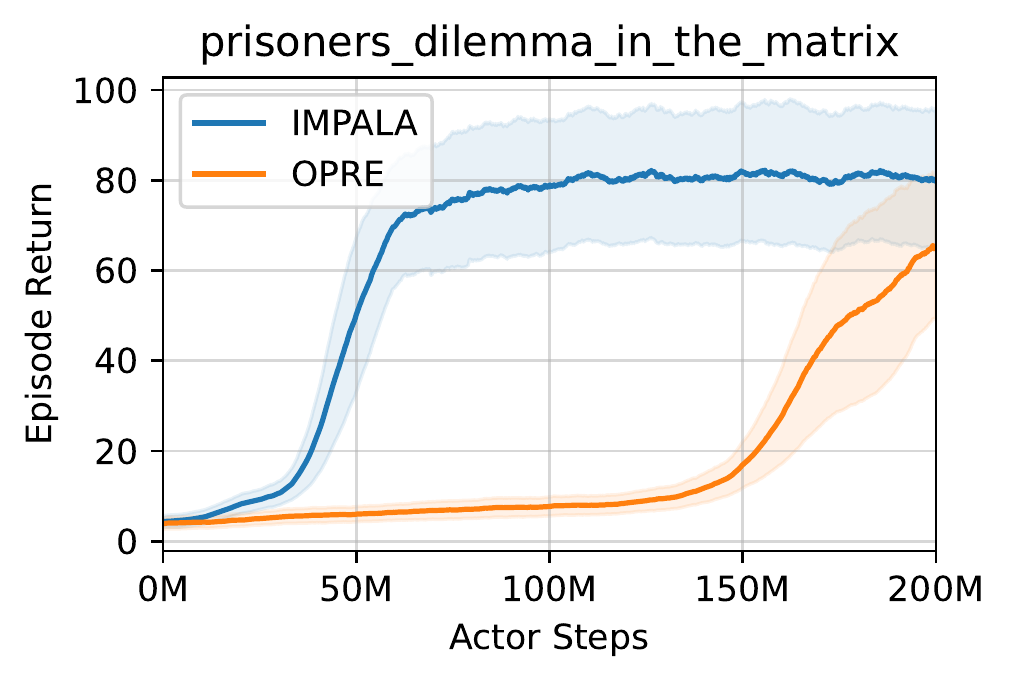}
    \caption{Training Plot on Prisoners Dilemma in the Matrix}
    \label{fig:dilemma}
\end{figure}

We evaluate our implementation in two environments to assess its performance and generalization capabilities across different types of multi-agent scenarios. The first environment, Meltingpot, encompasses a wide range of game types, including cooperative, competitive, and general-sum games. In Meltingpot, agents interact with each other to achieve various objectives, which can involve cooperation, competition, or a combination of both. This environment allows us to examine how well our implementation handles different types of interactions and strategies, evaluating its performance in cooperative, competitive, and general-sum settings.

The second environment, Overcooked, focuses specifically on cooperative multi-agent scenarios. In Overcooked, agents work together in a shared kitchen to prepare meals and serve customers. The emphasis in this environment is on effective coordination, communication, and cooperation among the agents to maximize efficiency and customer satisfaction. By evaluating our implementation in Overcooked, we can specifically assess its performance and effectiveness in cooperative multi-agent settings, where collaboration and teamwork are crucial for success.

By evaluating our implementation in both Meltingpot and Overcooked, we gain a comprehensive understanding of its performance in a range of multi-agent scenarios. This evaluation enables us to analyze how well our approach adapts to different types of interactions, strategies, and objectives, and provides valuable insights into its strengths and limitations. The findings from these evaluations contribute to advancing our understanding of multi-agent reinforcement learning and inform further research and development in this field.

\subsection{MeltingPot}
We conduct evaluations on four distinct environments from the Meltingpot-v1 and Meltingpot-v2 domains.

In Meltingpot-v1, we evaluate our approach on two environments:
\begin{itemize}
    \item \textbf{Running with Scissors in the Matrix}: The training progress is visualized in Figure~\ref{fig:rws}, and the evaluation scores across different episodes are provided in Table~\ref{tab:rws}.
    \item \textbf{Prisoners' Dilemma in the Matrix}: The training progress is depicted in Figure~\ref{fig:dilemma}, and the evaluation scores for various scenarios are presented in Table~\ref{tab:dilemma}.
\end{itemize}

\begin{figure}
    \centering
    \includegraphics[width=0.7\linewidth]{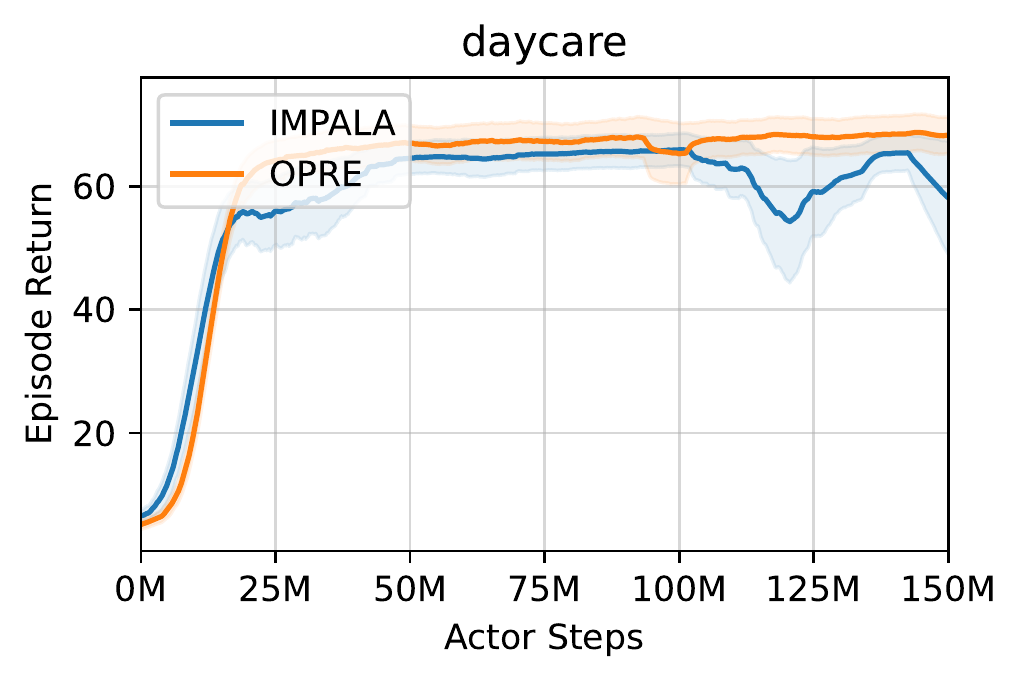}
    \caption{Training Plot on Daycare}
    \label{fig:daycare}
\end{figure}

\begin{figure}
    \centering
    \includegraphics[width=0.7\linewidth]{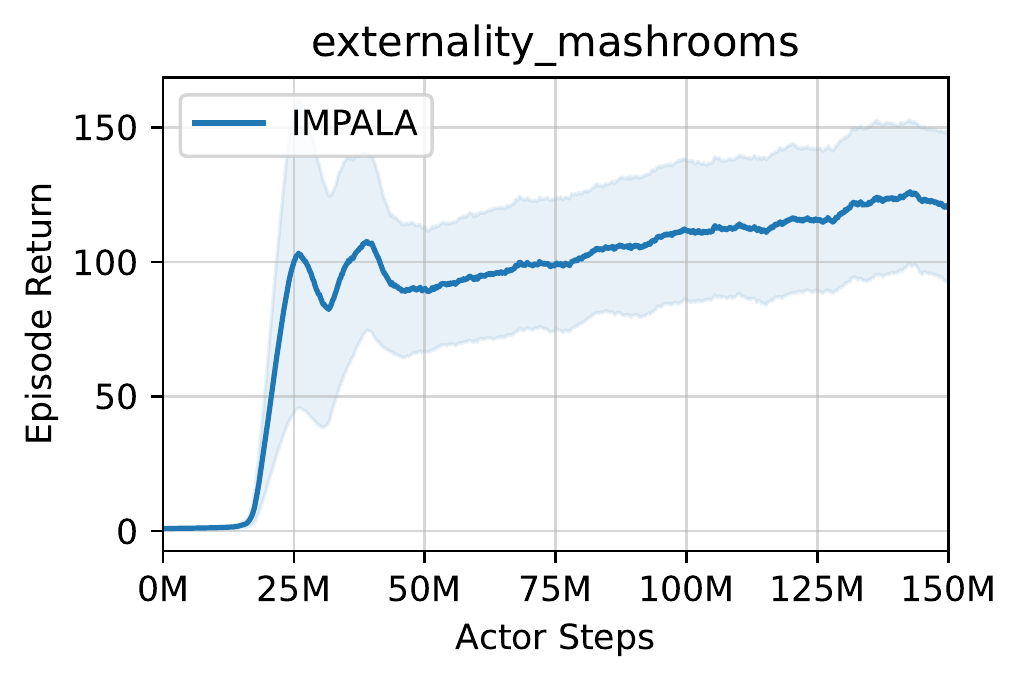}
    \caption{Training Plot on Externality Mushrooms}
    \label{fig:externality_mushrooms}
\end{figure}

In Meltingpot-v2, we assess our approach on two additional environments:
\begin{itemize}
    \item \textbf{Daycare}: The training progress is shown in Figure~\ref{fig:daycare}, and the evaluation scores for different scenarios are summarized in Table~\ref{tab:daycare}.
    \item \textbf{Externality Mushrooms Dense}: The training progress is illustrated in Figure~\ref{fig:externality_mushrooms}, and the evaluation scores for various scenarios are provided in Table~\ref{tab:externality_mushrooms}.
\end{itemize}
These evaluations allow us to analyze the performance of our approach across different environments and scenarios within the Meltingpot framework. The training plots provide insights into the learning progress, while the evaluation scores offer quantitative measures of the agent's performance in various scenarios.

\begin{table}[ht]
   \begin{center}  
    \begin{tabular}{|c|c|c|c|}
    \hline
     & \bfseries IMPALA & \bfseries OPRE \\
    \hline
    \bfseries Substrate & 65.94 & 67.83 \\
    \bfseries Scenario 0 & 0.89 & 0.33 \\
    \bfseries Scenario 1 & 109.11 & 126.00 \\
    \bfseries Scenario 2 & 0.22 & 0.00 \\
    \bfseries Scenario 3 & 154.56 & 171.33 \\
    \hline
    \end{tabular}
    \caption{Evaluation on Daycare}
    \label{tab:daycare}
    \end{center}
\end{table}

\begin{table}[ht]
   \begin{center}  
    \begin{tabular}{|c|c|c|c|}
    \hline
     & \bfseries IMPALA \\
    \hline
    \bfseries Scenario 0 & 547.70 \\
    \bfseries Scenario 1 & 13.21 \\
    \bfseries Scenario 2 & 293.02 \\
    \bfseries Scenario 3 & 38.04 \\
    \bfseries Substrate & 91.56 \\
    \hline
    \end{tabular}
    \caption{Evaluation on Externality Mushrooms}
    \label{tab:externality_mushrooms}
    \end{center}
\end{table}

\subsection{Overcooked}
\begin{figure}
    \centering
    \includegraphics[width=0.7\linewidth]{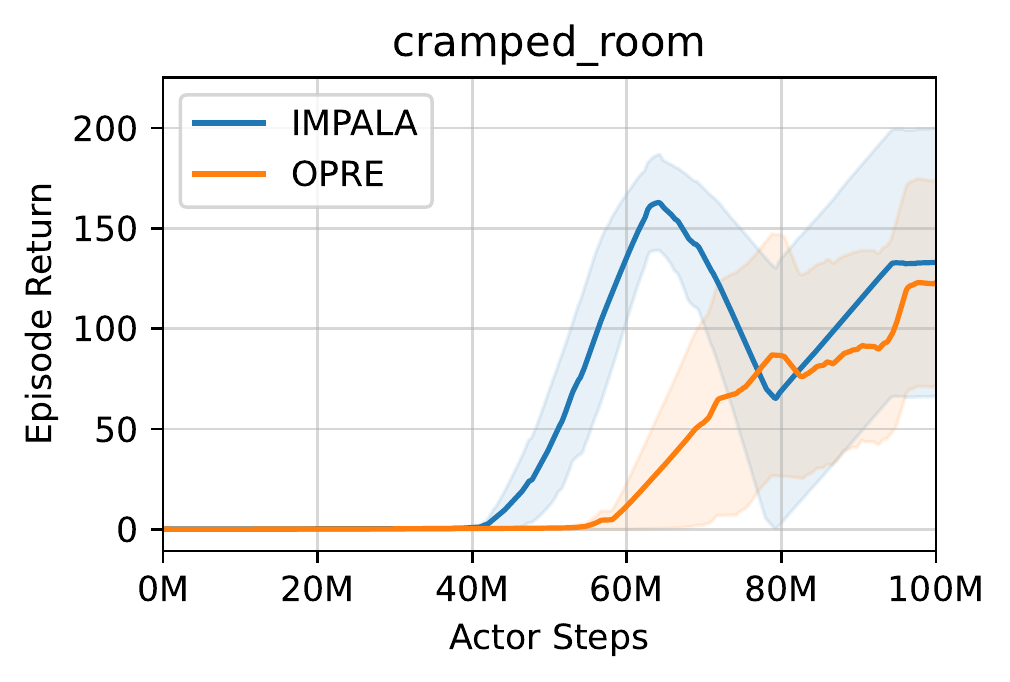}
    \caption{Training Plot on Cramped Room}
    \label{fig:cramped_room}
\end{figure}

We evaluate the performance of the algorithms on the \emph{Cramped Room} environment from the Overcooked domain. The training progress of the algorithms is visualized in Figure~\ref{fig:cramped_room}, providing insights into their learning dynamics and convergence behavior.

\section{Conclusion and Future Works}
In this paper, we introduced \libname, a highly optimized package for training and evaluation of the generalization of a population of agents to novel partners. The package provides an easy-to-use utility to train and evaluate the trained agents. It also provides an open-source implementation of ORPE~\cite{vezhnevetsOptionsResponsesGrounding2020}, a MARL algorithm designed for generalization. This package is targeted for researchers working on generalization in MARL and reduces the entry barrier for new researchers in MARL generalization. As any software package, \libname~is under continuous development and, in future, aims to implement other population learning algorithms \cite{openendedlearningteam2021openended}.

\bibliography{sample}

\begin{thebibliography}{27}
\providecommand{\natexlab}[1]{#1}
\providecommand{\url}[1]{\texttt{#1}}
\expandafter\ifx\csname urlstyle\endcsname\relax
  \providecommand{\doi}[1]{doi: #1}\else
  \providecommand{\doi}{doi: \begingroup \urlstyle{rm}\Url}\fi

\bibitem[Agapiou et~al.(2022)Agapiou, Vezhnevets, {Du{\'e}{\~n}ez-Guzm{\'a}n},
  Matyas, Mao, Sunehag, K{\"o}ster, Madhushani, Kopparapu, Comanescu, Strouse,
  Johanson, Singh, Haas, Mordatch, Mobbs, and Leibo]{agapiouMeltingPot2022}
John~P. Agapiou, Alexander~Sasha Vezhnevets, Edgar~A.
  {Du{\'e}{\~n}ez-Guzm{\'a}n}, Jayd Matyas, Yiran Mao, Peter Sunehag, Raphael
  K{\"o}ster, Udari Madhushani, Kavya Kopparapu, Ramona Comanescu, D.~J.
  Strouse, Michael~B. Johanson, Sukhdeep Singh, Julia Haas, Igor Mordatch, Dean
  Mobbs, and Joel~Z. Leibo.
\newblock Melting {{Pot}} 2.0, 2022.

\bibitem[Babuschkin et~al.(2020)Babuschkin, Baumli, Bell, Bhupatiraju, Bruce,
  Buchlovsky, Budden, Cai, Clark, Danihelka, Fantacci, Godwin, Jones, Hemsley,
  Hennigan, Hessel, Hou, Kapturowski, Keck, Kemaev, King, Kunesch, Martens,
  Merzic, Mikulik, Norman, Quan, Papamakarios, Ring, Ruiz, Sanchez, Schneider,
  Sezener, Spencer, Srinivasan, Wang, Stokowiec, and Viola]{deepmind2020jax}
Igor Babuschkin, Kate Baumli, Alison Bell, Surya Bhupatiraju, Jake Bruce, Peter
  Buchlovsky, David Budden, Trevor Cai, Aidan Clark, Ivo Danihelka, Claudio
  Fantacci, Jonathan Godwin, Chris Jones, Ross Hemsley, Tom Hennigan, Matteo
  Hessel, Shaobo Hou, Steven Kapturowski, Thomas Keck, Iurii Kemaev, Michael
  King, Markus Kunesch, Lena Martens, Hamza Merzic, Vladimir Mikulik, Tamara
  Norman, John Quan, George Papamakarios, Roman Ring, Francisco Ruiz, Alvaro
  Sanchez, Rosalia Schneider, Eren Sezener, Stephen Spencer, Srivatsan
  Srinivasan, Luyu Wang, Wojciech Stokowiec, and Fabio Viola.
\newblock The {D}eep{M}ind {JAX} {E}cosystem, 2020.

\bibitem[Bradbury et~al.(2018)Bradbury, Frostig, Hawkins, Johnson, Leary,
  Maclaurin, Necula, Paszke, Vander{P}las, Wanderman-{M}ilne, and
  Zhang]{jax2018github}
James Bradbury, Roy Frostig, Peter Hawkins, Matthew~James Johnson, Chris Leary,
  Dougal Maclaurin, George Necula, Adam Paszke, Jake Vander{P}las, Skye
  Wanderman-{M}ilne, and Qiao Zhang.
\newblock {JAX}: composable transformations of {P}ython+{N}um{P}y programs,
  2018.

\bibitem[Brockman et~al.(2016)Brockman, Cheung, Pettersson, Schneider,
  Schulman, Tang, and Zaremba]{2016openaigym}
Greg Brockman, Vicki Cheung, Ludwig Pettersson, Jonas Schneider, John Schulman,
  Jie Tang, and Wojciech Zaremba.
\newblock Openai gym, 2016.

\bibitem[Carroll et~al.(2019)Carroll, Shah, Ho, Griffiths, Seshia, Abbeel, and
  Dragan]{carrollUtilityLearningHumans2019overcooked}
Micah Carroll, Rohin Shah, Mark~K Ho, Tom Griffiths, Sanjit Seshia, Pieter
  Abbeel, and Anca Dragan.
\newblock On the {{Utility}} of {{Learning}} about {{Humans}} for {{Human-AI
  Coordination}}.
\newblock In \emph{Advances in {{Neural Information Processing Systems}}},
  volume~32. {Curran Associates, Inc.}, 2019.

\bibitem[Cassirer et~al.(2021)Cassirer, {Barth-Maron}, Brevdo, Ramos, Boyd,
  Sottiaux, and Kroiss]{cassirerReverbFrameworkExperience2021}
Albin Cassirer, Gabriel {Barth-Maron}, Eugene Brevdo, Sabela Ramos, Toby Boyd,
  Thibault Sottiaux, and Manuel Kroiss.
\newblock Reverb: {{A Framework For Experience Replay}}, 2021.

\bibitem[Castro et~al.(2018)Castro, Moitra, Gelada, Kumar, and
  Bellemare]{castro18dopamine}
Pablo~Samuel Castro, Subhodeep Moitra, Carles Gelada, Saurabh Kumar, and
  Marc~G. Bellemare.
\newblock Dopamine: {A} {R}esearch {F}ramework for {D}eep {R}einforcement
  {L}earning.
\newblock 2018.

\bibitem[Ellis et~al.(2022)Ellis, Moalla, Samvelyan, Sun, Mahajan, Foerster,
  and Whiteson]{ellis2022smacv2}
Benjamin Ellis, Skander Moalla, Mikayel Samvelyan, Mingfei Sun, Anuj Mahajan,
  Jakob~N Foerster, and Shimon Whiteson.
\newblock Smacv2: An improved benchmark for cooperative multi-agent
  reinforcement learning.
\newblock \emph{arXiv preprint arXiv:2212.07489}, 2022.

\bibitem[Espeholt et~al.(2018)Espeholt, Soyer, Munos, Simonyan, Mnih, Ward,
  Doron, Firoiu, Harley, Dunning, Legg, and
  Kavukcuoglu]{espeholtIMPALAScalableDistributed2018}
Lasse Espeholt, Hubert Soyer, Remi Munos, Karen Simonyan, Vlad Mnih, Tom Ward,
  Yotam Doron, Vlad Firoiu, Tim Harley, Iain Dunning, Shane Legg, and Koray
  Kavukcuoglu.
\newblock {{IMPALA}}: {{Scalable Distributed Deep-RL}} with {{Importance
  Weighted Actor-Learner Architectures}}.
\newblock In \emph{Proceedings of the 35th {{International Conference}} on
  {{Machine Learning}}}. {PMLR}, 2018.

\bibitem[Espeholt et~al.(2019)Espeholt, Marinier, Stanczyk, Wang, and
  Michalski]{espeholt2019seed}
Lasse Espeholt, Rapha{\"e}l Marinier, Piotr Stanczyk, Ke~Wang, and Marcin
  Michalski.
\newblock Seed rl: Scalable and efficient deep-rl with accelerated central
  inference.
\newblock 2019.

\bibitem[Hessel et~al.(2021)Hessel, Kroiss, Clark, Kemaev, Quan, Keck, Viola,
  and van Hasselt]{hessel2021podracer}
Matteo Hessel, Manuel Kroiss, Aidan Clark, Iurii Kemaev, John Quan, Thomas
  Keck, Fabio Viola, and Hado van Hasselt.
\newblock Podracer architectures for scalable reinforcement learning.
\newblock 2021.

\bibitem[Hoffman et~al.(2022)Hoffman, Shahriari, Aslanides, {Barth-Maron},
  Momchev, Sinopalnikov, Sta{\'n}czyk, Ramos, Raichuk, Vincent, Hussenot,
  Dadashi, {Dulac-Arnold}, Orsini, Jacq, Ferret, Vieillard, Ghasemipour,
  Girgin, Pietquin, Behbahani, Norman, Abdolmaleki, Cassirer, Yang, Baumli,
  Henderson, Friesen, Haroun, Novikov, Colmenarejo, Cabi, Gulcehre, Paine,
  Srinivasan, Cowie, Wang, Piot, and {de
  Freitas}]{hoffmanAcmeResearchFramework2022}
Matthew~W. Hoffman, Bobak Shahriari, John Aslanides, Gabriel {Barth-Maron},
  Nikola Momchev, Danila Sinopalnikov, Piotr Sta{\'n}czyk, Sabela Ramos, Anton
  Raichuk, Damien Vincent, L{\'e}onard Hussenot, Robert Dadashi, Gabriel
  {Dulac-Arnold}, Manu Orsini, Alexis Jacq, Johan Ferret, Nino Vieillard, Seyed
  Kamyar~Seyed Ghasemipour, Sertan Girgin, Olivier Pietquin, Feryal Behbahani,
  Tamara Norman, Abbas Abdolmaleki, Albin Cassirer, Fan Yang, Kate Baumli,
  Sarah Henderson, Abe Friesen, Ruba Haroun, Alex Novikov, Sergio~G{\'o}mez
  Colmenarejo, Serkan Cabi, Caglar Gulcehre, Tom~Le Paine, Srivatsan
  Srinivasan, Andrew Cowie, Ziyu Wang, Bilal Piot, and Nando {de Freitas}.
\newblock Acme: {{A Research Framework}} for {{Distributed Reinforcement
  Learning}}, 2022.

\bibitem[Hu et~al.(2022)Hu, Zhong, Gao, Wang, Dong, Li, Liang, Chang, and
  Yang]{hu2022marllib}
Siyi Hu, Yifan Zhong, Minquan Gao, Weixun Wang, Hao Dong, Zhihui Li, Xiaodan
  Liang, Xiaojun Chang, and Yaodong Yang.
\newblock Marllib: A scalable multi-agent reinforcement learning library.
\newblock \emph{arXiv preprint arXiv:2210.13708}, 2022.

\bibitem[Huang et~al.(2022)Huang, Dossa, Ye, Braga, Chakraborty, Mehta, and
  Ara{\'u}jo]{huangCleanRLHighqualitySinglefile2022}
Shengyi Huang, Rousslan Fernand~Julien Dossa, Chang Ye, Jeff Braga, Dipam
  Chakraborty, Kinal Mehta, and Jo{\~a}o G.~M. Ara{\'u}jo.
\newblock {{CleanRL}}: {{High-quality Single-file Implementations}} of {{Deep
  Reinforcement Learning Algorithms}}.
\newblock \emph{Journal of Machine Learning Research}, 23\penalty0 (274), 2022.
\newblock ISSN 1533-7928.

\bibitem[Liang et~al.(2018)Liang, Liaw, Nishihara, Moritz, Fox, Goldberg,
  Gonzalez, Jordan, and Stoica]{liangRLlibAbstractionsDistributed2018}
Eric Liang, Richard Liaw, Robert Nishihara, Philipp Moritz, Roy Fox, Ken
  Goldberg, Joseph Gonzalez, Michael Jordan, and Ion Stoica.
\newblock {{RLlib}}: {{Abstractions}} for {{Distributed Reinforcement
  Learning}}.
\newblock In \emph{Proceedings of the 35th {{International Conference}} on
  {{Machine Learning}}}. {PMLR}, 2018.

\bibitem[Mahajan et~al.(2022)Mahajan, Samvelyan, Gupta, Ellis, Sun,
  Rockt{\"a}schel, and Whiteson]{mahajan2022generalization}
Anuj Mahajan, Mikayel Samvelyan, Tarun Gupta, Benjamin Ellis, Mingfei Sun, Tim
  Rockt{\"a}schel, and Shimon Whiteson.
\newblock Generalization in cooperative multi-agent systems.
\newblock \emph{arXiv preprint arXiv:2202.00104}, 2022.

\bibitem[Muldal et~al.(2019)Muldal, Doron, Aslanides, Harley, Ward, and
  Liu]{dm_env2019}
Alistair Muldal, Yotam Doron, John Aslanides, Tim Harley, Tom Ward, and Siqi
  Liu.
\newblock dm\_env: A python interface for reinforcement learning environments,
  2019.

\bibitem[Papoudakis et~al.(2021)Papoudakis, Christianos, Schäfer, and
  Albrecht]{papoudakis2021epymarl}
Georgios Papoudakis, Filippos Christianos, Lukas Schäfer, and Stefano~V.
  Albrecht.
\newblock Benchmarking multi-agent deep reinforcement learning algorithms in
  cooperative tasks.
\newblock In \emph{Proceedings of the Neural Information Processing Systems
  Track on Datasets and Benchmarks (NeurIPS)}, 2021.

\bibitem[Pretorius et~al.(2021)Pretorius, Tessera, Smit, Formanek, Grimbly,
  Eloff, Danisa, Francis, Shock, Kamper, Brink, Engelbrecht, Laterre, and
  Beguir]{pretoriusMavaResearchFramework2021}
Arnu Pretorius, Kale-ab Tessera, Andries~P. Smit, Claude Formanek, St~John
  Grimbly, Kevin Eloff, Siphelele Danisa, Lawrence Francis, Jonathan Shock,
  Herman Kamper, Willie Brink, Herman Engelbrecht, Alexandre Laterre, and Karim
  Beguir.
\newblock Mava: A research framework for distributed multi-agent reinforcement
  learning, 2021.

\bibitem[Raffin et~al.(2021)Raffin, Hill, Gleave, Kanervisto, Ernestus, and
  Dormann]{raffinStableBaselines3ReliableReinforcement2021}
Antonin Raffin, Ashley Hill, Adam Gleave, Anssi Kanervisto, Maximilian
  Ernestus, and Noah Dormann.
\newblock Stable-{{Baselines3}}: {{Reliable Reinforcement Learning
  Implementations}}.
\newblock \emph{Journal of Machine Learning Research}, 22\penalty0 (268), 2021.
\newblock ISSN 1533-7928.

\bibitem[Samvelyan et~al.(2019)Samvelyan, Rashid, de~Witt, Farquhar, Nardelli,
  Rudner, Hung, Torr, Foerster, and Whiteson]{samvelyan19smac}
Mikayel Samvelyan, Tabish Rashid, Christian~Schroeder de~Witt, Gregory
  Farquhar, Nantas Nardelli, Tim G.~J. Rudner, Chia-Man Hung, Philiph H.~S.
  Torr, Jakob Foerster, and Shimon Whiteson.
\newblock {The} {StarCraft} {Multi}-{Agent} {Challenge}.
\newblock \emph{CoRR}, abs/1902.04043, 2019.

\bibitem[Sarkar et~al.(2022)Sarkar, Talati, Shih, and
  Sadigh]{sarkarPantheonRLMARLLibrary2022}
Bidipta Sarkar, Aditi Talati, Andy Shih, and Dorsa Sadigh.
\newblock {{PantheonRL}}: {{A MARL Library}} for {{Dynamic Training
  Interactions}}.
\newblock \emph{Proceedings of the AAAI Conference on Artificial Intelligence},
  36\penalty0 (11), 2022.
\newblock ISSN 2374-3468.

\bibitem[Team et~al.(2021)Team, Stooke, Mahajan, Barros, Deck, Bauer,
  Sygnowski, Trebacz, Jaderberg, Mathieu, McAleese, Bradley-Schmieg, Wong,
  Porcel, Raileanu, Hughes-Fitt, Dalibard, and
  Czarnecki]{openendedlearningteam2021openended}
Open Ended~Learning Team, Adam Stooke, Anuj Mahajan, Catarina Barros, Charlie
  Deck, Jakob Bauer, Jakub Sygnowski, Maja Trebacz, Max Jaderberg, Michael
  Mathieu, Nat McAleese, Nathalie Bradley-Schmieg, Nathaniel Wong, Nicolas
  Porcel, Roberta Raileanu, Steph Hughes-Fitt, Valentin Dalibard, and
  Wojciech~Marian Czarnecki.
\newblock Open-ended learning leads to generally capable agents, 2021.

\bibitem[Terry et~al.(2021)Terry, Black, Grammel, Jayakumar, Hari, Sullivan,
  Santos, Dieffendahl, Horsch, Perez-Vicente, et~al.]{terry2021pettingzoo}
J~Terry, Benjamin Black, Nathaniel Grammel, Mario Jayakumar, Ananth Hari, Ryan
  Sullivan, Luis~S Santos, Clemens Dieffendahl, Caroline Horsch, Rodrigo
  Perez-Vicente, et~al.
\newblock Pettingzoo: Gym for multi-agent reinforcement learning.
\newblock \emph{Advances in Neural Information Processing Systems}, 2021.

\bibitem[Vezhnevets et~al.(2020)Vezhnevets, Wu, Leblond, and
  Leibo]{vezhnevetsOptionsResponsesGrounding2020}
Alexander~Sasha Vezhnevets, Yuhuai Wu, Remi Leblond, and Joel~Z. Leibo.
\newblock Options as responses: {{Grounding}} behavioural hierarchies in
  multi-agent {{RL}}, 2020.
\newblock URL \url{http://arxiv.org/abs/1906.01470}.

\bibitem[Yang et~al.(2021)Yang, {Barth-Maron}, Sta{\'n}czyk, Hoffman, Liu,
  Kroiss, Pope, and Rrustemi]{yangLaunchpadProgrammingModel2021}
Fan Yang, Gabriel {Barth-Maron}, Piotr Sta{\'n}czyk, Matthew Hoffman, Siqi Liu,
  Manuel Kroiss, Aedan Pope, and Alban Rrustemi.
\newblock Launchpad: {{A Programming Model}} for {{Distributed Machine Learning
  Research}}, 2021.

\bibitem[Zhou et~al.(2023)Zhou, Wan, Wang, Wen, Wu, Wen, Yang, Yu, Wang, and
  Zhang]{malib}
Ming Zhou, Ziyu Wan, Hanjing Wang, Muning Wen, Runzhe Wu, Ying Wen, Yaodong
  Yang, Yong Yu, Jun Wang, and Weinan Zhang.
\newblock Malib: A parallel framework for population-based multi-agent
  reinforcement learning.
\newblock \emph{Journal of Machine Learning Research}, 24\penalty0
  (150):\penalty0 1--12, 2023.
\newblock URL \url{http://jmlr.org/papers/v24/22-0169.html}.

\end{thebibliography}

\end{document}